\documentclass[times,tight]{aastex62}

\usepackage{amsfonts}
\usepackage{mathrsfs}

\def\cblue{\color{black}}

\def\sunm{M_{\odot}}

\newcommand{\BHm}{M_{\bullet}}
\newcommand{\bhm}{m_{\bullet}}

\newcommand{\calR}{\dot{\cal R}_{\rm Bondi}}

\newcommand{\dotMBon}{\dot{M}_{\rm Bon}}
\newcommand{\ergs}{{\rm erg\,s^{-1}}}
\newcommand{\RBon}{R_{\rm Bon}}
\newcommand{\Rg}{R_{\rm g}}

\newcommand{\mathdotM}{\dot{\mathscr{M}}}

\newcommand\kms{\rm km\,s^{-1}}

\received{\it 2020 Feburary 11}
\revised{\it 2021 March 10, 2021}
\accepted{\it 2021 March 12, 2021}

\shorttitle{Accretion-modified Stars in AGNs}
\shortauthors{Wang et al.}

\begin{document}

\title{\large\bf Accretion-modified Stars in Accretion Disks of Active Galactic Nuclei:
Slowly Transient Appearance}

\author[0000-0001-9449-9268]{Jian-Min Wang}
\affil{Key Laboratory for Particle Astrophysics, Institute of High Energy Physics,
Chinese Academy of Sciences, 19B Yuquan Road, Beijing 100049, China}
\affil{School of Astronomy and Space Sciences, University of Chinese Academy of Sciences, 
19A Yuquan road, Beijing 100049, China}
\affil{National Astronomical Observatory of China, 20A Datun Road, Beijing 100020, China}

\author{Jun-Rong Liu}
\affil{Key Laboratory for Particle Astrophysics, Institute of High Energy Physics,
Chinese Academy of Sciences, 19B Yuquan Road, Beijing 100049, China}
\affil{School of Astronomy and Space Sciences, University of Chinese Academy of Sciences, 
19A Yuquan road, Beijing 100049, China}

\author{Luis C. Ho}
\affil{Kavli Institute for Astronomy and Astrophysics, Peking University, Beijing 100871, China}
\affil{Department of Astronomy, School of Physics, Peking University, Beijing 100871, China}

\author{Pu Du}
\affil{Key Laboratory for Particle Astrophysics, Institute of High Energy Physics,
Chinese Academy of Sciences, 19B Yuquan Road, Beijing 100049, China}

\begin{abstract}
Compact objects are expected to exist in the accretion disks of supermassive black holes (SMBHs) 
in active galactic nuclei (AGNs), and in the presence of such a dense environment 
($\sim 10^{14}\,{\rm cm^{-3}}$), they will form a new kind of stellar population denoted as 
Accretion-Modified Stars (AMSs). This hypothesis is supported by recent LIGO/Virgo detection 
of the mergers of very high-mass stellar binary black holes (BHs). We show that the AMSs will be trapped 
by the SMBH-disk within a typical AGN lifetime. In the context of SMBH-disks, the rates of Bondi accretion 
onto BHs are $\sim 10^{9}L_{\rm Edd}/c^{2}$, where $L_{\rm Edd}$ is the Eddington luminosity and $c$ is 
the speed of light. Outflows developed from the hyper-Eddington accretion strongly impact the Bondi sphere 
and induce episodic accretion. We show that the hyper-Eddington accretion will be halted after an accretion
interval of $t_{\rm a}\sim 10^{5}m_{1}\,$s, where $m_{1}=\bhm/10\sunm$ is the BH mass. The kinetic energy 
of the outflows accumulated during $t_{\rm a}$ is equivalent to 10 supernovae driving an explosion of the 
Bondi sphere and developing blast waves. We demonstrate that a synchrotron flare from relativistic electrons
accelerated by the blast waves peaks in the soft X-ray band ($\sim 0.1\,$keV), significantly contributing 
to the radio, optical, UV, and soft X-ray emission of typical radio-quiet quasars. External inverse Compton
scattering of the electrons peaks around $40\,$GeV and is detectable through {\it Fermi}-LAT. The flare,
decaying with $t^{-6/5}$ with a few months, will appear as a slowly varying transient. The flares, 
occurring at a rate of a few per year in radio-quiet quasars, provide a new mechanism for explaining 
 AGN variability. 
\end{abstract}
\subjectheadings{Active galactic nuclei (16); Galaxy accretion disks (562); Supermassive black holes (1663)}

\section{Introduction}
Compact objects (neutron stars and stellar black holes) may exist in accretion disks 
of active galactic nuclei (AGNs) and quasars. On the one hand, pioneering ideas of 
self-gravitating accretion disks in AGNs \citep{Paczynski1977,Kolykhalov1980,Shlosman1989} 
have received increasing 
attention \citep{Collin1999,Collin2008,Goodman2003,Goodman2004,Wang2010,Wang2011,Wang2012}, 
for they offer a possible explanation for the super-solar metallicities inferred in AGNs and 
quasars \citep{Hamann1998,Warner2003,Nagao2006,Shin2013,Du2014}. The high metallicities can 
be naturally linked to star formation in self-gravitating disks, which inevitably produces
compact objects from supernova explosions. On the other hand, the high metallicity of quasars 
can  also be explained by stars from nuclear clusters captured by SMBH-disks \citep{Artymowicz1993}, 
a process that will also introduce compact objects in the disks \citep{Cheng1999,Cantiello2020}. 
Enveloped by the very dense gaseous medium of the accretion disk, compact objects inevitably will 
form a special kind of object, whose envelope is very massive, even more massive that the 
central compact object itself (see \S2.2). We use the terminology of Accretion-Modified Stars
(AMSs) to denote these special objects in the SMBH-disks\footnote{The original concept of the 
Thorne-\.Zytkow objects (TZOs: \citealt{Thorne1975,Thorne1977}), which only pertained to neutron 
stars enshrouded by envelopes are in both hydrostatic and 
thermal equilibrium, with some accretion occurring just in the very inner region close to the 
neutron stars. They are expected to have specific surface abundances due to exotic nuclear 
burning processes happening deeper into the star and close to the compact object. The present
AMSs are neither in hydrostatic nor thermal equilibrium. These compact objects are episodically  
accreting from the Bondi sphere. }. 
The fate of AMSs in such an environment, however, is unknown.

The recent detection of GW190521, consisting of $(85+66)\sunm$ binary BHs, by the Advanced 
LIGO/Virgo consortium \citep{Abbott2020} has received much attention because it is harbored 
by the quasar SDSS J1249+3449 monitored by the Zwicky Transient Facility \citep{Graham2020}. 
BH binaries with such high masses far exceeds the limit of pair instability of massive stars 
\citep[e.g.,][]{Woosley2017}. As the BH masses of GW190521 are much
 higher than the upper limit of supernova explosions from an isolated star,
this event has inspired the idea that mergers of compact objects can occur in AGN disks 
\citep{Cheng1999,Bartos2017,McKernan2019,Yang2019,Yang2020,Tanaga2020,Samsing2020}. 
Such a scenario would involve AMSs in SMBH-disks. 

In this paper, we suggest that BHs are undergoing episodic accretion governed by powerful 
outflows developed from hyper-Eddington accretion. The outflows drive a Bondi explosion and result 
in non-thermal flares from the blast waves that are predicted to appear as slowly varying transients 
in the radio, optical, UV, and soft X-ray bands. AMSs offer a new mechanism for explaining AGN 
variability but also for origins of massive stellar black holes.

\begin{figure*}
    \centering
    \includegraphics[width= 0.8\textwidth,trim=-15 170 10 95,clip]{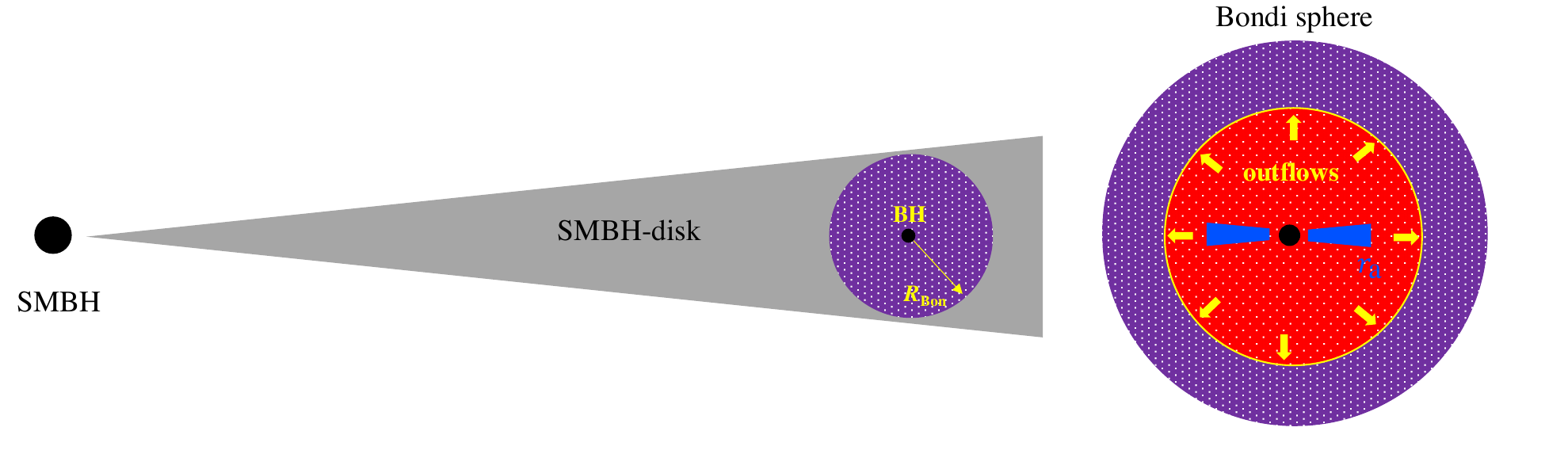}
    \caption{{\it Left}: Stellar mass black holes ($1\sunm-10^{2}\sunm$) form Accretion-modified
    Stars (AMSs) in AGN accretion disks. {\it Right}: The Bondi sphere hosts a BH undergoing 
    hyper-Eddington accretion. Powerful outflows developed from the slim disk of
    the BH produce strong feedback to the Bondi sphere. The hyper-accretion is halted by the
    mechanical feedback at the radius $r_{\rm a}(\sim 10^{4}$ gravitational radii) within 
    interval $t_{\rm a}$, generating flares from radio to $\gamma$-rays (see text for details).
    In this paper, we neglect the spins of the Bondi sphere (see a brief discussion in \S2.2).
    }
\end{figure*}

\section{A simple model}
Stars, which originate either from capture from nuclear star clusters or direct formation 
from self-gravitating disks, accrete gas from SMBH-disks to form more massive objects that 
evolve quickly into compact objects \citep{Artymowicz1993,Cheng1999,Cantiello2020}. 
The massive envelope around a compact object forms an AMS through accretion of gas from the local 
SMBH-disk. In this paper, we only focus on AMSs of black holes.  Other more complicated options 
can arise. 

The dimensionless accretion rate of the central SMBH is defined by 
$\mathdotM=\dot{M}_{\bullet}/\dot{M}_{\rm Edd}$, where $\dot{M}_{\rm Edd}=L_{\rm Edd}c^{-2}$ 
is the Eddington limit rate, $\dot{M}_{\bullet}$ is the accretion rate of the SMBH, 
$L_{\rm Edd}=4\pi G\BHm m_{\rm p}c/\sigma_{\rm T}=1.3\times10^{46}M_{8}\,{\rm erg\,s^{-1}}$,
$G$ is the gravitational constant, $M_{8}=\BHm/10^{8}\sunm$ is the SMBH mass in units of
$10^{8}\sunm$, $c$ is the speed of light, $m_{\rm p}$ is the proton mass, and $\sigma_{\rm T}$
is the Thomson cross section.  
The half-thickness, density, mid-plane temperature, and radial velocity of the SMBH-disk are 
\citep[e.g.,][]{Kato2008}
\begin{equation}\label{Eq:disk}
\left\{
\begin{array}{l}\vspace{1ex}
H=4.3\times 10^{14}\,\alpha_{0.1}^{-1/10}M_{8}^{9/10}\mathdotM^{3/20} r_{4}^{9/8}\,{\rm cm},\\ \vspace{1ex}
\rho_{\rm d} =6.9\times 10^{-11}\,\left(\alpha_{0.1}M_{8}\right)^{-7/10}\mathdotM^{11/20}r_{4}^{-15/8}\,{\rm g\,cm^{-3}},\\ \vspace{1ex}
T_{c}=4.6\times10^{3}\,\left(\alpha_{0.1}M_{8}\right)^{-1/5}\mathdotM^{3/10}r_{4}^{-3/4}\,{\rm K},\\
v_{r}=2.6\times 10^{2}\,\alpha_{0.1}^{4/5}M_{8}^{-1/5}\mathdotM^{3/10}r_{4}^{-1/4}\,{\rm cm\,s^{-1}},
\end{array}\right.
\end{equation}
where $\alpha_{0.1}=\alpha/0.1$ is the viscosity parameter, $r_{4}=R/10^{4}\Rg$ is the radius 
of the disk from the SMBH, and $\Rg=G\BHm/c^{2}$ is the gravitational radius. The self-gravity 
of the disk can be described by the Toomre parameter, defined as 
$Q=\Omega_{\rm K} c_{s}/\pi G \rho_{\rm d} H$,
where $c_{s}\approx \sqrt{3kT_{c}/m_{p}}\approx 15.7\,T_{4}^{1/2}\,\kms$ is the local sound speed of
SMBH-disk and $\Omega_{\rm K}=\sqrt{G\BHm/R^{3}}$. The disk becomes self-gravitating (SG) beyond a 
critical radius where $Q=1$, which is given by
\begin{equation}
R_{\rm SG}/\Rg=1.2\times 10^{3}\,\alpha_{0.1}^{28/45}M_{8}^{-52/45}\mathdotM^{-22/45}.
\end{equation}
We consider the regions beyond $R_{\rm SG}$, where massive stars are formed either through 
capture by the SMBH-disk or they are formed in situ, likely with a top-heavy initial mass 
function because of the high temperatures in the nuclear environment.

We would like to point out that the self-gravitating regions are undergoing more 
complicated physics beyond the solutions of Eqn. (\ref{Eq:disk}).  For example, star formation 
is unavoidable \citep{Collin1999,Collin2008,Sirko2003,Thompson2005}, gravito turbulence 
dominates \citep{Rafikov2007,Rafikov2015}, and the region has  a clumpy distribution. The 
actual density and temperature could be different from that used here so that 
the fate of AMSs could be more complicated than currently assumed. The main characteristics 
of their evolution, however, can be predicted qualitatively.
Future studies will be carried out about compact objects colliding with gaseous clumps 
in the self-gravitating regions.

\subsection{AMSs}
Considering that black holes formed from massive stars are kicked off with random velocities of
$v_{\bullet}^{\prime}\approx 10^{2}-10^{3}\,\kms$ with respect to their remnant (the black holes
are still tightly bound by the SMBHs), accretion onto the black holes in SMBH-disks can be 
described by the Hoyle-Lyttleton-Bondi (HLB) formulation. 
%
%
Neglecting the weak dependence on 
adiabatic index and self-gravity of the accreted gas \citep{Wandel1984}, we take the simplest 
form
\begin{equation}
\dotMBon=\frac{4\pi G^{2}\bhm^{2}\rho_{\rm d}}{\left(c_{s}^{2}+v_{\bullet}^{2}\right)^{3/2}},
\end{equation}
where $\bhm$ is the black hole mass from supernova explosion and $v_{\bullet}$ is the kick-off 
velocity relative to the motion of the SMBH-disk. The kick-off velocity of the black holes is 
uncertain, but it could be up to $v_{\bullet}^{\prime}\sim 10^{3}\,\kms$ if comparable 
with the case of neutron stars \citep[e.g.,][]{Nakamura2019}. We consider two extreme scenarios, 
in which the black hole is co-rotating and counterrotating with the SMBH-disk.
If $v_{\bullet}^{\prime}\sim V_{\rm K}=3000\,r_{4}^{-1/2}\,\kms$, with $V_{\rm K}$ the Keplerian 
velocity, we expect two populations of accreting black holes in the disk: co-rotating black holes 
with $v_{\bullet}\approx |V_{\rm K}-v_{\bullet}^{\prime}|\ll c_{s}$ and counterrotating ones with 
$v_{\bullet}\approx V_{\rm K}+v_{\bullet}^{\prime}\approx 2V_{\rm K}\gg c_{s}$
(see \S\ref{Sec:friction}). It should be stressed that the current choice of kick-off 
velocities are just for the two possible regimes, and is not necessarily representative of the 
bulk population. The corresponding accretion rates are
\begin{equation}\label{Eq:dotm}
\dot{m}_{\rm Bon}=\frac{\dotMBon}{\dot{M}_{\rm Edd}}\approx\left\{\begin{array}{ll}\vspace{1ex}
8.9\times 10^{8}\,m_{1}(\alpha_{0.1}M_{8})^{-2/5}\mathdotM^{1/10}r_{4}^{-3/4}&({\textrm{for\, corotation})}, \\
5.1\,m_{1}(\alpha_{0.1}M_{8})^{-7/10}\mathdotM^{11/20}r_{4}^{-3/8} &({\textrm{for\, counter\, rotation}}),
\end{array}\right.
\end{equation}
and the Bondi radius $R_{\rm Bon}=G\bhm/\left(c_s^{2}+v_{\bullet}^{2}\right)$,
\begin{equation}
\RBon=\left\{\begin{array}{ll}\vspace{1ex}
1.2\times 10^{15}\,m_{1}\left(\alpha_{0.1}M_{8}\right)^{1/5}\mathdotM^{-3/10}r_{4}^{3/4}\,{\rm cm}
&({\textrm{for\, corotation})},\\
3.7\times 10^{9}\,m_{1}r_{4}\,{\rm cm} &({\textrm{for\,counter\, rotation}}),
\end{array}\right.
\end{equation}
where $m_{1}=\bhm/10\sunm$ is the mass of the black hole inside the SMBH-disk. The Hill 
radius is 
$R_{\rm Hill}=(M_{\rm Bon}/\BHm)^{1/3}R
  \approx 1.5\times 10^{15}\,M_{\rm Bon,2}^{1/3}M_{8}^{2/3}r_{4}\,{\rm cm}$, 
which constrains the size of the AMS. Since $\RBon\approx R_{\rm Hill}$, tidal disruption of the AMS
can be avoided.
The validity of the HLB accretion is guaranteed by  $\RBon\ll R_{\rho}$, where
$R_{\rho}=\left|d\ln\rho_{\rm d}/dR\right|^{-1}$ is the density scale of the SMBH-disk
(i.e., the length that the density changes by an $e$-folding factor). 
A schematic illustrating an AMS embedded in a SMBH-disk is shown in the left panel of Figure 1.

It should be noted that such a hyper-Eddington accretion is much higher than the case 
in the early Universe, where seed black holes only attain a dimensionless accretion 
rate of $\dot{m}_{\bullet}\sim$ a few hundred \citep{Volonteri2005,Toyouchi2020,Takeo2020}. 
Radiative feedback to the seed growth dominates in this case
\citep{Wang2006,Milos2009a,Milos2009b,Regan2019}. It depends on the Compton 
temperature, which in turn relies on the hard X-ray spectral slope. 
The slopes get steeper with increases of accretion rates \citep{Wang2003,Wang2004}, 
lowering the Compton temperature. Moreover, luminosity saturation of the slim disk decreases 
the relative emission in hard X-rays \citep{Abramowicz1988,Wang1999}. Considering this fact, 
radiative feedback 
becomes weaker with increases of accretion rates. Radiative feedback could be still important 
for $\dot{m}_{\bullet}\sim 10^{3}$ from numerical simulations \citep{Takeo2020}, but mechanical 
feedback of outflows could dominate if $\dot{m}_{\bullet}\sim 10^{9}$ discussed in the current 
cases.

\subsection{Friction on AMSs}\label{Sec:friction}
The Bondi mass, which is defined as gas mass within the Bondi radius, can be approximated
by $M_{\rm Bon}\approx 4\pi R_{\rm Bon}^{3}\rho_{\rm d}/3$
\begin{equation}
M_{\rm Bon}=\left\{\begin{array}{ll}\vspace{1ex}
2.4\times 10^{2}\sunm\,m_{1}^{3}(\alpha_{0.1}M_{8})^{-1/10}\mathdotM^{-7/20}r_{4}^{3/8}&{\textrm{(for co-rotation)}},\\
7.3\times 10^{-15}\sunm\,m_{1}^{3}(\alpha_{0.1}M_{8})^{-7/10}\mathdotM^{11/20}r_{4}^{9/8}&{\textrm{(for counter rotation)}.}
\end{array}\right.
\end{equation} 
We note that the co-rotating AMSs are much massive than those counter rotating. 
According to Equation (2) in \cite{Artymowicz1993}, the drag force on an AMS parallel to its path 
is given by $F_{\rm d}=4\pi G^{2}M_{\rm AMS}^{2}\rho_{\rm d}v_{\bullet}^{-2}$
for $v_{\bullet}<v_{\rm esc}$, where $v_{\rm esc}=\sqrt{2GM_{\rm AMS}/R_{\rm Bon}}$
is escaping velocity from the AMS surface, where $M_{\rm AMS}=M_{\rm Bon}+\bhm\approx (M_{\rm Bon},\bhm)$
for co-rotating and counterrotating AMSs, respectively. Therefore, the slow-down timescale is given by
$F_{\rm d}\times (v_{\bullet} t_{\rm slow})=M_{\rm AMS}v_{\bullet}^{2}/2$, we have
\begin{equation}\label{Eq:slow}
t_{\rm slow}=\left\{\begin{array}{ll}\vspace{1ex}
14.2\,M_{\rm Bon,2}^{-1}\rho_{\bar{10}}^{-1}v_{100}^{3}\,{\textrm{yr}}&\textrm{(for co-rotation)},\\
3.8\times 10^{6}\,m_{1}^{-1}\rho_{\bar{10}}^{-1}v_{3000}^{3}\,{\textrm{yr}}&\textrm{(for counter rotation),}
\end{array}\right.
\end{equation}
where $v_{100,3000}=v_{\bullet}/(100,3000)\,\kms$, 
$\rho_{\bar{10}}=\rho_{\rm d}/10^{-10}\,{\rm{g\,cm^{-3}}}$. 
We find that the counterrotating AMSs need much longer times to slow-down than the co-rotating ones,
and they have to spend a significant fraction of the typical AGN lifetime ($t_{\rm AGN}
\sim R/v_{r}\sim 10^{7}\,r_{4}v_{2}^{-1}$yr, where $v_{2}=v_{r}/10^{2}\,{\rm cm\,s^{-1}}$ from Eq.1).
The randomly kicked BHs may leave the SMBH-disk for some cases if $v_{\bullet}t_{\rm slow}\gtrsim H$, 
but they are still bound by the SMBH so that they are finally trapped by the disks. As fully random
kick-off from supernova explosion, black holes, whose velocities perpendicular to co-rotation 
direction will be dramatically damped according to Eq.(\ref{Eq:slow}), and finally trapped by the
SMBH-disks.

Finally, we point out that the spin and angular momentum distribution of the AMSs 
are complicated issues. On the one hand, the accreted gas spins up the AMSs with an opposite 
direction to rotation of the SMBH-disk since the AMS side near to the SMBH is rotating faster than 
its outside, namely AMS spins are from differential rotation of the disk. On the other hand, 
the tidal torque of the SMBH is still sufficiently strong to drive the AMS spin (at the 
characteristic radius $10^{4}R_{\rm g}$), 
although the AMSs avoid the tidal disruption. The tidal torque drives the Bondi sphere to co-rotate 
with the disk when the Bondi sphere is large enough ({\cblue like the case of the moon, whose 
spin follows its orbit around the earth through the 
tidal torque of the earth}). Therefore, the outer part of the Bondi sphere generally has low spin 
because of the action of the negative tidal torque, however, 
the inner part may have some angular momentum where the torque is insufficiently strong. This could be 
the reason why there are slim accretion disks around the central black holes in the AMSs. This needs 
some detailed calculations for the AMSs as done for trapped stars in the SMBH-disks by \cite{Jermyn2021}, 
but it is beyond the scope of this paper.

\subsection{Hyper-Eddington accretion onto black holes}
As shown by Equation (\ref{Eq:dotm}), the black holes of AMSs have an extremely high accretion rate 
($\sim 10^{9}\,\dot{M}_{\rm Edd}$), which have never been discussed in literatures (except for accretion
of neutrons onto black holes for $\gamma$-ray bursts). There 
are two schools of models of super-Eddington accretion onto black holes: 1) the classical
model is the so-called slim disks without outflows \citep{Abramowicz1988,Wang1999} and 2) 
accretion onto black holes with strong outflows \citep[e.g.,][]{Ohsuga2005}. Actual physics 
could be between the two schools: photon trapping and outflows co-exist \citep{Kitaki2018}. 
The first generally shows a cold disk with strong photon trapping. Simulations show that 
super-Eddington accretion produces powerful outflows strongly influencing its surroundings,
but the mid-plane still continues accretion with rates from $300$ to $10^{3}L_{\rm Edd}/c^{2}$.
The current accretion rates are higher than the cases discussed in \cite{Takeo2020} by 5 orders. 
We introduce two parameters in the BH accretion: 1) $f_{\rm a}$ as a fraction of the 
Bondi rates contributing to BH growth \citep{Takeo2020}
and 2) $f_{\rm out}(\sim 1)$ as a fraction of the Bondi rates to outflows. We
have $\dot{m}_{\rm grow}=f_{\rm a}\dot{m}_{\rm Bon}$.

The hyper-Eddington accreting black hole of the AMS has never been studied so far. In  principle
it should be an self-consistent system under the government of accretion and (radiative and kinematic) 
feedback. When accretion rates are intermediately super-Eddington compared with the current context, 
such as $10^{3}L_{\rm Edd}/c^{2}$, radiative feedback to its surroundings dominates and drives the
accretion to be episodic \citep{Wang2006,Milos2009a,Milos2009b}. If the hyper-Eddington accretion
develops very powerful outflows with kinematic luminosity ($L_{\rm kin}$), for example, with a power 
of $\sim \dot{m}L_{\rm Edd}$, the kinematic momentum-driven feedback will dominate. The swept shell 
of the outflows follows \citep{King2003}
\begin{equation}\label{Eq:momentum}
\frac{d}{dt}\left(M_{r}\dot{r}\right)=\frac{L_{\rm kin}}{v_{\rm out}},
\end{equation}
where $M_{r}$ is the shell mass inside $r$ and $v_{\rm out}$ is the velocity of the outflows.
The kinematic luminosity during a single accretion episode is given by
\begin{equation}\label{Eq:Lkin}
L_{\rm kin}=\eta f_{\rm out}\dot{m}_{\rm Bon}L_{\rm Edd}=1.3\times 10^{47}\,\eta_{0.1}m_{1}\dot{m}_{9}\,
 {\rm erg\,s^{-1}},
\end{equation}
where $\dot{m}_{9}=\dot{m}_{\rm Bon}/10^{9}$, $\eta$ is the dissipating efficiency rather than 
the radiative efficiency and determined by the last stable orbit around black holes. Usually 
$\eta=0.1$ is taken in literatures. The post-shock gas has a temperature
$kT_{\rm sh}\approx 9m_{\rm p}v_{\rm out}^{2}/16$,
and accretion onto the BH is halted when $T_{\rm sh}$ is higher than the virial temperature
$kT_{\rm vir}\approx m_{\rm p}c^{2}(r/r_{\rm g})^{-1}$. We have the condition 
\begin{equation}\label{Eq:vout}
v_{\rm out}=\frac{4}{3}\frac{c}{\sqrt{r_{\rm a}/r_{\rm g}}},
\end{equation}
to quench the accretion onto the black hole at a radius $r_{\rm a}$ of the slim disks. 
Integrating Equation (\ref{Eq:momentum}), we have
\begin{equation}\label{Eq:momentum2}
M_{r}v_{\rm out}\approx \left(\frac{L_{\rm kin}}{v_{\rm out}}\right)t_{\rm a},
\end{equation}
where $t_{\rm a}$ is the accretion timescale. Here, we neglect the initial condition in the
integration. Since the radial velocity of the hyper-Eddington accretion is 
$v_{r}=\alpha V_{\rm K}/\sqrt{5}$
\citep[see Eq. 11 in][]{Wang1999}, 
the accretion may be halted beyond $r_{\rm a}$ after 
\begin{equation}\label{Eq:tacc}
t_{\rm a}=\frac{r_{\rm a}}{v_{r}}.
\end{equation}
Combining Equations (\ref{Eq:Lkin}-\ref{Eq:tacc}), we have
\begin{equation}
\frac{r_{\rm a}}{r_{\rm g}}
  =\left(\frac{16\alpha}{9\sqrt{5}}\frac{M_{\rm Bon}c^{3}}{r_{\rm g}L_{\rm kin}}\right)^{2/5}
  =1.3\times 10^{5}\,\eta_{0.1}^{-2/5}\alpha_{0.1}^{13/25}M_{8}^{3/25}\mathdotM^{-9/50}r_{4}^{9/20}
\end{equation}
and
\begin{equation}
t_{\rm a}=\left(\frac{16\sqrt[3]{5}}{9}
          \frac{c^{4/3}r_{\rm g}^{2/3}M_{\rm Bon}}{\alpha^{2/3}L_{\rm kin}}\right)^{3/5}
           =5.1\times 10^{4}\,m_{1}\eta_{0.1}^{-3/5}\alpha_{0.1}^{-11/50}M_{8}^{9/50}
           \mathdotM^{-27/100}r_{4}^{27/40}\,{\textrm{s}},
\end{equation}
where we take $M_{r}=M_{\rm Bon}$. The results show the necessary accretion timescale to
quench the Bondi accretion through the outflows from slim accretion disks. $t_{\rm a}$ and
$r_{\rm a}$ are not sensitive to the density distribution of the Bondi sphere. 
Fig.1 right panel shows a cartoon of the Bondi explosion driven by the powerful outflows
from slim accretion disks with hyper-Eddington rates.

\subsection{Bondi explosion}
The momentum-driven outflows are able to push the Bondi sphere within $t_{\rm a}$, but the
cumulative kinematic energies during the period is much larger than the self-gravitational energy
of the Bondi sphere so that it is undergoing explosion driven by cumulative energy of the
outflows
($E_{\rm SG}\approx GM_{\rm AMS}^{2}/R_{\rm Bon}\approx 10^{48}M_{\rm AMS,2}^{2}R_{\rm Bon,15}^{-1}\,$erg
$\ll\, E_{\rm kin}$). Fig.2 shows a cartoon of the Bondi explosion from a SMBH-disk to 
the BLR. The cumulative kinematic energies within the sphere is
\begin{equation}
E_{\rm kin}=L_{\rm kin}t_{\rm a}=1.3\times 10^{52}\,\eta_{0.1}m_{1}\dot{m}_{9}t_{\rm a,5}\,{\rm erg},
\end{equation}
where $t_{\rm a,5}=t_{\rm a}/10^{5}\,{\rm s}$. We denote this as the Bondi explosion. Energies 
of one Bondi explosion are equivalent to that of about 10 supernovae.  The explosion in the BLR (with medium density 
$\rho_{\rm BLR}$) as a quasi-sphere can be approximately 
described by the Sedov self-solution. However, the SMBH-disk is not sphere, we use the adiabatic 
approximation for the expansion in the disk. Taking the SMBH-disk as a slab, we have its opening 
solid angle to the Bondi sphere is about $\epsilon_{\rm d}\approx 0.5\left(H/R_{\rm exp}\right)^{2}$, 
where $R_{\rm exp}$ is the expanding radius of the Bondi explosion in the SMBH-disk. Using the Sedov 
solution in BLR and the adiabatic approximation of 
$4\pi R_{\rm exp}^{2}Hn_{\rm d}kT_{\rm d}=\epsilon_{\rm d} E_{\rm kin}$ in the SMBH-disk
\begin{equation}\label{Eq:Rexp}
R_{\rm exp}=\left\{\begin{array}{ll}\vspace{1ex}
    \displaystyle{\left(\frac{E_{\rm kin}}{\rho_{\rm BLR}}\right)^{1/5}t^{2/5}}
  =55.1\,E_{52}^{1/5}n_{\rm BLR,7}^{-1/5}t_{10}^{2/5}\,{\textrm{ltd}}& (\textrm{for BLR medium}),\\
\displaystyle{\left(\frac{HE_{\rm kin}}{8\pi n_{\rm d}kT_{\rm d}}\right)^{1/4}}
  =9.0\times 10^{15}\,E_{52}^{1/4}\alpha_{0.1}^{1/5}M_{8}^{9/20}\mathdotM^{-7/40}r_{4}^{15/16}\,{\rm cm}
     &(\textrm{for SMBH-disk}),
\end{array}\right.
\end{equation}
the explosion velocity is
\begin{equation}
V_{\rm exp}=\left\{\begin{array}{ll}\vspace{1ex}
\displaystyle{\frac{2}{5}\left(\frac{E_{\rm kin}}{\rho_{\rm BLR}}\right)^{1/5}t^{-3/5}}
=1.8\times 10^{3}\,E_{52}^{1/5}n_{\rm BLR,7}^{-1/5}t_{10}^{-3/5}\,\rm{km\,s^{-1}}& 
(\textrm{for BLR medium}),\\
\displaystyle{
   \left(\frac{2HL_{\rm kin}}{\pi n_{\rm d}kT_{\rm d}t_{a}^{3}}\right)^{1/4}
  =1.7\times 10^{6}\,\alpha_{0.1}^{1/5}M_{8}^{9/20}
  \mathdotM^{-7/40}r_{4}^{15/16}L_{47}^{1/4}t_{\rm a,5}^{-3/4}\,\rm{km\,s^{-1}}}&(\textrm{for SMBH-disk}),
\end{array}\right.
\end{equation}
from $4\pi R_{\rm exp}H(n_{\rm d}kT_{\rm d})V_{\rm exp}=\epsilon_{\rm d}L_{\rm kin}$ for SMBH-disk,
and the explosion timescale is
\begin{equation}
t_{\rm exp}=\left\{\begin{array}{ll}\vspace{1ex}
\displaystyle{\left(\frac{3}{4\pi}\right)^{5/6}
             \frac{M_{\rm Bon}^{5/6}}{\rho_{\rm BLR}^{1/3}E_{\rm kin}^{1/2}}}
  =9.9\,M_{\rm Bon,2}^{5/6}n_{\rm BLR,7}^{-1/3}E_{52}^{-1/2}\,{\textrm{yr}}&(\textrm{for BLR medium}),\\
 t_{\rm a}/2& (\textrm{for SMBH-disk}),
\end{array}\right.
\end{equation}
from the condition of $M_{\rm Bon}=4\pi R_{\rm exp}^{2}V_{\rm exp}\rho_{\rm BLR}t_{\rm exp}$, and 
$t_{\rm exp}=R_{\rm exp}/V_{\rm exp}$ for the SMBH-disk, where 
$n_{\rm BLR,7}=n_{\rm BLR}/10^{7}\,{\rm cm^{-3}}$, $n_{\rm BLR}=\rho_{\rm BLR}/m_{\rm p}$, $E_{52}=E_{\rm kin}/10^{52}\,{\rm erg}$,
$L_{47}=L_{\rm kin}/10^{47}\,{\rm erg\,s^{-1}}$, $t_{10}=t/10\,{\rm yr}$, 
$n_{14}=n_{\rm d}/10^{14}\,{\rm cm^{-3}}$ and $n_{\rm d}=\rho_{\rm d}/m_{\rm p}$ 
is the number density of the SMBH-disk. The temperature of the shock-swept medium is
\begin{equation}
T\approx \frac{2(\Gamma_{\rm ad}-1)m_{\rm p}}{(1+\Gamma_{\rm ad})^{2}k}V_{\rm exp}^{2}
 =2.3\times 10^{7}\,V_{\rm exp,3}^{2}\,{\textrm{K}},
\end{equation}
where $V_{\rm exp,3}=V_{\rm exp}/10^{3}\,{\rm{km\,s^{-1}}}$ and we take the adiabatic index 
$\Gamma_{\rm ad}=5/3$. It should be noted that the expansion velocity of the SMBH-disk is 
relativistic and equivalent to a Lorentz factor $\Gamma_{\rm exp}\approx V_{\rm exp}/c\approx 5$. 
A cavity with a radius $R_{\rm exp}$ is formed by the Bondi explosion in the SMBH-disk. The small 
fraction $\epsilon_{\rm d}(\sim 10\%)$ of $E_{\rm kin}$ could be thermalized in the SMBH-disk, 
we leave the relativistic blast waves as an open topic in future. {\cblue The Bondi explosion 
expands into the BLR medium.}

\begin{figure*}
    \centering
    \includegraphics[width= 0.64\textwidth,trim=-15 -20 10 5,clip]{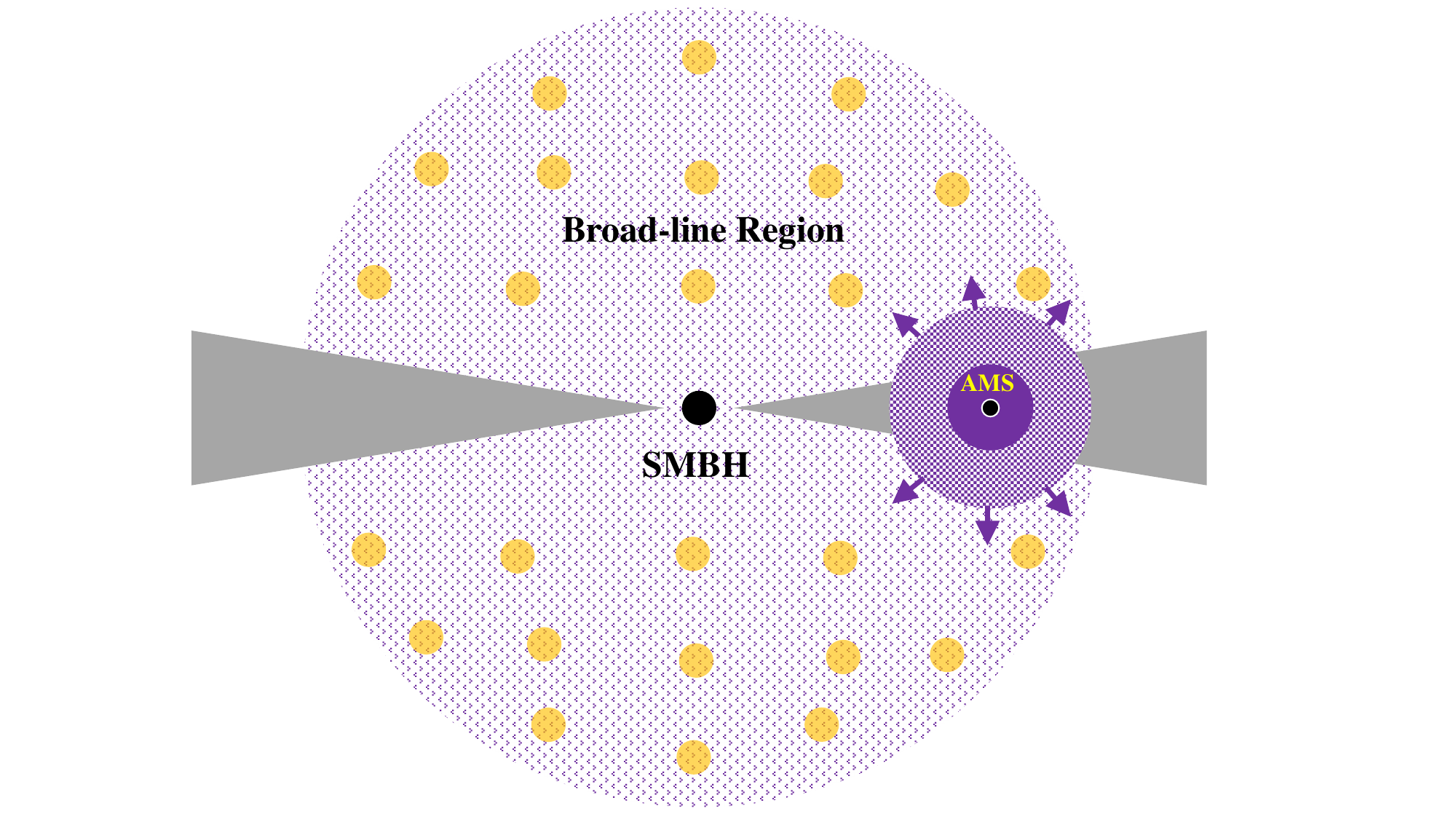}
    \caption{The Bondi explosion driven by the hyper-Eddington accretion onto the black hole of 
    the AMS. The plot is shown on a logarithmic scale. The explosion 
    forms a cavity in the SMBH-disk. The yellow balls are clouds in the broad-line region. The 
    explosion propagates into the broad-line region with a density lower than that of the SMBH-disk 
    by a factor of $10^{7}$, and thus it extends much beyond the disk. Such a violent explosion, 
    equivalent to $\sim$10 supernovae, results in a slowly varying transient in many bands.}
\end{figure*}

Here we would like to stress that the above scenario of the Bondi explosion is the most 
conservative. The Bondi spheres are actually streaming toward the central black holes,
leading to the possibility that the shocks can be enhanced through collision with the outflow.
If we include radial self-gravity of the Bondi sphere \citep[e.g.,][]{Wandel1984}, the shocks 
could be enhanced further. The streaming kinetic energy could be of the same order of magnitude 
as $E_{\rm SG}$, therefore, the characteristic features of the 
explosion remains. By the way, it is interesting to compare the long $\gamma$-ray bursts (GRBs) 
and the Bondi explosion for similarity and difference between them. \cite{Woosley1993} pioneered 
an idea about the long GRBs. They originate from failed type Ib supernovae of massive stars, however,
highly relativistic jets are developed from disk accretion of neutrons onto the central black holes 
with neutrinos cooling. Their accretion rates, 
$\sim 1\,\sunm\,{\rm s^{-1}}\approx 10^{15}\dot{M}_{\rm Edd}$, are much higher than the AMS cases. 
The similarity is that both kinds 
of explosions are driven by accretion onto black holes, but the differences rest on that not only 
cooling mechanisms are distinguished but also the GRBs are more violent in the extremely compact 
regions than the Bondi explosion. Unlike the GRBs dominated by relativistic jets,
the Bondi explosions are driven by powerful outflows appearing as slow transients.

\subsection{Rejuvenation of AMSs}
The cavity formed by the Bondi explosion makes the BH to have very low accretion rates, and thus
they are hardly detectable individually. However, the cold gas of the SMBH-disk replenishes the 
cavity rejuvenating AMSs in a timescale of 
\begin{equation}
t_{\rm rej}=\frac{R_{\rm exp}}{c_{s}}
   =268.9\,\alpha_{0.1}^{3/10}M_{8}^{11/20}\mathdotM^{-13/40}r_{4}^{21/16}E_{52}^{1/4}\,{\rm yr}.
\end{equation}
The duty cycle of the BH accretion is 
$\delta_{\bullet}=t_{\rm a}/(t_{\rm a}+t_{\rm rej})\approx 6\times 10^{-6}$.
The episodic accretion efficiently constrains on the BH growth. With the duty cycle,
growth timescale is about
$t_{\rm grow}=(\delta_{\bullet}\dot{m}f_{\rm a})^{-1}t_{\rm Salp}
            =10\,(\delta_{\bar{5}}\dot{m}_{9}f_{\rm a,\bar{3}})^{-1}\,t_{\rm Salp}$, 
where $t_{\rm Salp}=\bhm/\dot{M}_{\rm Edd}=0.45\,$Gyr is the Salpeter time, 
$\delta_{\bar{5}}=\delta_{\bullet}/10^{-5}$, $f_{\rm a,\bar{3}}=f_{\rm a}/10^{-3}$ is the fraction 
of the outflow to the accretion rates. The most uncertainties are $f_{\rm a}$, but it is 
easy for the BHs to grow up to $10^{2}\sunm$ from $10\sunm$. This needs numerical simulations 
of hyper-Eddington accretion at small scale of $\lesssim 10^{3}r_{\rm g}$.

It should be noted that most BHs are quiescent because of the extremely low duty 
cycles in SMBH-disks. If they are in sub-Eddington accretion status, they are radiating 
in X-ray bands like X-binaries. However, emissions from this kind of AMSs (with low accretion 
rates) depend on details of their surroundings and their mass functions. Total emissions of 
these accreting BHs may significantly contribute to the observed. It would be an interesting 
topic in future.

\section{Observational Signatures of AMSs}
\subsection{Emissions from the Bondi explosion}
The Bondi explosion can be divided into two phases. First, internal shocks due to the 
collision between the outflows and the Bondi sphere. The dissipated energies could be of 
the orders of $E_{\rm SG}$ and channelled into thermal energy because of the very large
optical depth $\tau_{\rm es}^{0}\approx \kappa_{\rm es}\rho_{\rm d} R\sim 10^{4}$ for initial 
Bondi sphere with $R_{\rm Bon}\sim 10^{15}\,$cm
and $\rho_{\rm d}\sim 10^{-10}\,{\rm g\,cm^{-3}}$, where $\kappa_{\rm es}$ is electron scattering
opacity. The optical depth of the expanding Bondi sphere is
$\tau_{\rm es}\approx \kappa_{\rm es}(3M_{\rm Bon}/4\pi R^{3})R\propto R^{-2}$, we find 
$\tau_{\rm es}=\tau_{\rm es}^{0}(R_{\rm Bon}/R)^{2}$. {\cblue Since the Bondi explosion is mostly 
in the BLR medium, we neglect the part in the SMBH-disk.} When 
$R\sim 10^{2}R_{\rm Bon}\sim 10^{17}\,$cm, this phase ends within an interval of 
$\delta t_{\tau_{\rm es}}\sim 10^{6}\,$s from Eq. (\ref{Eq:Rexp}). We then have a variation
of luminosity $\Delta L\sim E_{\rm SG}/\delta t_{\tau_{\rm es}}\approx 10^{42}\,{\rm erg\,s^{-1}}$.
This is is too weak to observe for a quasar.

When $R=100R_{\rm Bon}$, the explosion begins to non-thermally radiate. For the
simplest estimation, we assume that electrons accelerated by shocks will generate non-thermal 
emissions with a fraction about $\xi$ of $E_{\rm kin}$. A giant flare of non-thermal 
emission with a luminosity is about $\Delta L\propto V_{\rm exp}^{2}$, we have
\begin{equation}
\Delta L_{\rm AGN}\approx L_{0}\left(\frac{t}{t_{\rm es}}\right)^{-6/5}\,
           {(\textrm{for}~ t\ge t_{\rm es})},
\end{equation}
from the self-similar expansion in the SMBH-disk, where 
$L_{0}=\xi E_{\rm kin}/5t_{\rm es}=1.0\times 10^{44}\,\xi_{0.05}E_{52}t_{\rm es,6}^{-1}\,{\rm erg\,s^{-1}}$,
$\xi_{0.05}=\xi/0.05$ \citep[e.g.,][]{Blandford1987}, 
and $t_{\rm es,6}=t_{\rm es}/10^{6}\,{\rm s}$. $L_{0}$ is determined by the total energy of
$\xi E_{\rm kin}$.
Spectral energy distributions (SEDs) depend on relativistic electrons and surrounding photons 
(synchrotron radiation and inverse Compton scattering), but this 
is a significant contribution to the steady luminosity of quasars. 

\begin{figure*}\label{fig:sed-lc}
    \centering
    \includegraphics[width= 0.65\textwidth]{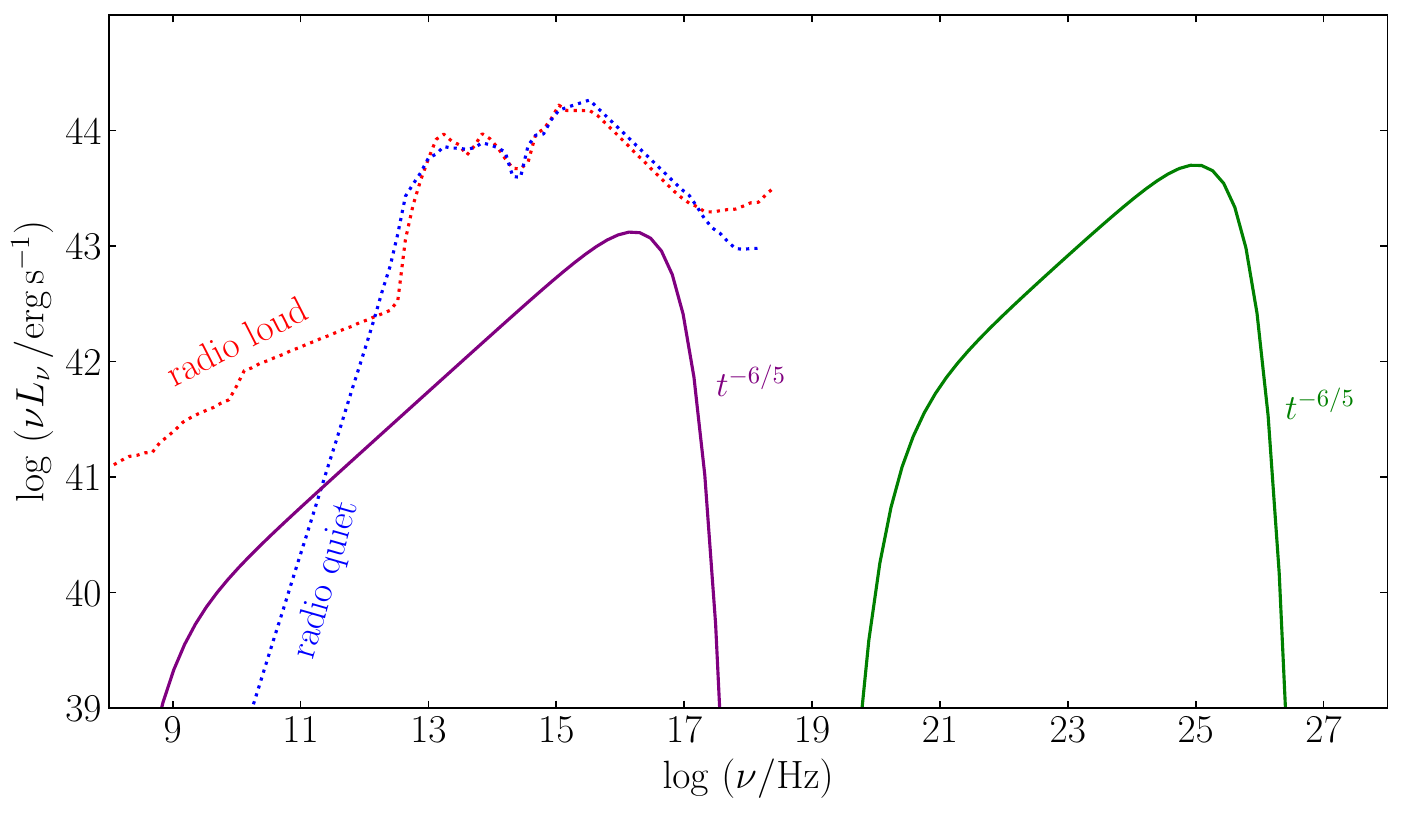}
    \caption{Spectral energy distributions (SEDs) of the Bondi explosion. Non-thermal emission
    from the explosion is characterized by SEDs peaking in the UV and TeV bands.
    Synchrotron emission is shown in purple, and external inverse Compton scattering is in green.
    The flare decays with time as $t^{-6/5}$, but the TeV photons are absorbed by 
    pair production. We show the mean SEDs \citep{Shang2011} of radio-quiet (dotted blue) and 
    radio-loud (dotted red) quasars for $\BHm=10^{8}\sunm$ and $\mathdotM=1$ (its bolometric 
    luminosity of $10^{45}\,{\rm erg\,s^{-1}}$). We scale the mean quasar SED to match 
    $L_{5100}\approx 0.1L_{\rm bol}$. The luminosities at the synchrotron and inverse Compton 
    peak frequency are $\sim 0.5(L_{\rm IC},L_{\rm syn})$ for relativistic electrons with $n_{e}=2$.}
\end{figure*}

In order to show the characteristic of light curves, we assume that the shocks accelerate electrons to
have a power law as $N_{\gamma}\propto \gamma^{-n_{e}}$ ($\gamma_{\rm min}\le \gamma\le\gamma_{\rm max}$) 
with an index $n_{e}$, where $\gamma$ is the Lorentz factor of relativistic electrons. The maximum 
energy of electron is determined by the balance between the acceleration and cooling. With the
equipartition with the BLR hot phase, we have a magnetic field of $B\approx 0.6\,(n_{7}T_{7})^{1/2}\,$G
from $u_{\rm B}=n_{\rm BLR}kT_{\rm hot}$, where $u_{\rm B}=B^{2}/8\pi$ is energy density of 
the magnetic field and $T_{\rm hot}=10^{7}T_{7}$ is the temperature of the hot phase of
the BLR. To illustrate the flare generated by the Bondi explosion, we consider a radio-quiet 
quasar with $\BHm=10^8{\sunm}$ and $\mathdotM=1$. Its bolometric luminosity is
$L_{\rm bol}=1.3\times 10^{45}\,\eta_{0.1}M_{8}\mathdotM\,{\rm erg\,s^{-1}}$.
Photon energy density in the BLR is about 
$u_{\rm ph}=L_{\rm bol}/4\pi R^{2}c\approx 0.16\,L_{45}R_{\rm 50}^{-2}\,{\rm erg\,cm^{-3}}$
much higher than the magnetic fields, where $L_{45}=L_{\rm bol}/10^{45}\ergs$, 
$R_{\rm 50}=R/{\rm 50\,ltd}$. This leads to external inverse Compton (IC) scattering as dominate 
cooling of the relativistic electrons accelerated by shocks. Acceleration timescale is given by
$t_{\rm acc}=R_{\rm L}c/V_{\rm exp}^{2}\approx 5.8\times 10^{2}\,\gamma_{5}B_{0}^{-1}T_{7}^{-1}\,$s
\citep{Blandford1987}, while the inverse Compton cooling timescale is 
$t_{\rm IC}=3m_{e}c/4\sigma_{\rm T}\gamma u_{\rm ph}=1.9\times 10^{3}\,\gamma_{5}^{-1}u_{0.16}^{-1}\,$s,
where $R_{\rm L}$ is the Larmor's radius, $\gamma_{5}=\gamma/10^{5}$ and $B_{0}=B/1\,{\rm G}$.
We have $\gamma_{\rm max}=1.8\times 10^{5}\,u_{0.16}^{-1/2}B_{0}^{1/2}T_{7}^{1/2}$ from 
$t_{\rm acc}=t_{\rm IC}$, and the maximum frequency of synchrotron and IC radiation are given by
\begin{equation}
\nu^{\rm max}_{\rm syn}\approx 0.17\,B_{0}\gamma_{\rm max,5}^{2}\,{\rm keV},\quad 
\nu^{\rm max}_{\rm IC}\approx \gamma_{\rm max}^{2}\nu_{\rm UV}
                      \approx 41.7\,\gamma_{\rm max,5}^{2}\nu_{15}^{\rm UV}\,{\rm GeV},
\end{equation}
with the peak luminosities of
\begin{equation}
L_{\rm syn}=2.5\times 10^{43}\,\left(\frac{u_{\rm B}/u_{\rm ph}}{0.25}\right)
             \left(\frac{L_{\rm IC}}{1\times 10^{44}{\rm erg\,s^{-1}}}\right){\rm erg\,s^{-1}},\quad
L_{\rm IC}\approx 1\times 10^{44}\,\xi_{0.05}E_{52}t_{\rm es,6}^{-1}\,{\rm erg\,s^{-1}},
\end{equation}
for an spectral index of $L_{\nu}\propto \nu^{(1-n_{e})/2}$, where 
$\nu_{15}^{\rm UV}=\nu_{\rm UV}/10^{15}\,{\rm Hz}$ is the UV photon frequency from the SMBH-disk. 
In Fig. 3, we show the mean SED of a typical quasar in order to
compare with multiwaveband light curves of the Bondi explosion.

From Fig.3, several remarkable features are found: 1) The synchrotron radiation peaks between 
soft X-rays (contributes to $1/3$) and has significant contribution to optical and UV (about 
$\sim 0.1$); 2) the Bondi explosion drives a radio-quiet quasar to 
become radio-intermediate in radio bands ($n_{e}=2$); 3) The external inverse Compton peaks at 
$\sim 40$\,GeV, and with about $5\times 10^{43}\,\ergs$
which can be detected in non-blazar AGNs by {\it Fermi}-LAT; 4) the transient appearance with
a temporal profile as $t^{-6/5}$ and decaying a factor 10 within about $6.8t_{\rm es}$, namely, 
3 months for $10^{8}\sunm$ quasars and the $10\sunm$ BH; 5) the intra-band emissions have no delays 
since they originate from the same regions. Moreover, for one BH with a few tens of solar 
mass, the above features will be more prominent. These characteristics are unique characteristics 
when identifying flares in AGNs. 
 
Expansion of the Bondi explosion exhausts a small fraction of the total energy ($\sim 10\%$) 
in the SMBH-disk, and most energy is released outside the disk as slow transients from radio
to $\gamma$-rays. This is very different from that of the relativistic expansion driven by jets
from neutron stars or black holes in SMBH-disks 
discussed in \cite{Perna2021} and \cite{Zhu2020} (actually for GRBs in the disks). Future 
detections of AGN light curves are useful to distinguish the nature of the AMSs in the SMBH-disks. 

\subsection{Bondi explosion rates}
According to the Kennicutt-Schmidt (KS) law of 
$\dot{\Sigma}_{*}=2.5\times 10^{-4}\Sigma_{0}^{1.4}\,\sunm\,{\rm yr^{-1}\,kpc^{-2}}$, star 
formation timescale is $t_{*}=\Sigma_{0}/\dot{\Sigma}_{*}=4\times 10^{7}\,\Sigma_{5}^{-0.4}\,$yr, 
where $\Sigma_{0,5}=\Sigma_{\rm gas}/(10^{0},10^{5})\,\sunm\,{\rm pc^{-2}}$ is gas surface density 
of star formation regions. We note that $t_{*}$ is comparable with $t_{\rm AGN}$.
From Eq. (1), we have surface density of SMBH-disks of 
$\Sigma_{\rm disk}=2.8\times 10^{8}\,\alpha_{0.1}^{-4/5}M_{8}^{1/5}\mathdotM^{7/10}r_4^{-3/4}\,
                       \sunm{\rm pc^{-2}}$,
the KS law results in a rate of $\sim 2.5\sunm\,{\rm yr}^{-1}$ in the SMBH-disk. This means that 
the KS star formation exhausts all disk gas in a timescale much shorter than  
$t_{\rm AGN}$ by a factor of $0.04$. This is obviously inconsistent with observations of AGN lifetime. 
One way to overcome this inconsistency is to decrease the star formation efficiency or only a small 
fractions of SMBH-disk mass converting into stars. Star formation in this region is poorly 
constrained by observations \citep[e.g.,][]{Collin2008}.  For concreteness, we assume that the 
maximum star formation is that all SMBH-disk gas is converted into stars with a rate of about 
$M_{\rm disk}/t_{\rm AGN}$, and black hole numbers are of 
$N_{\bullet}\sim M_{\rm disk}/\langle m_{*}\rangle$ over the AGN episode, where 
$M_{\rm disk}\approx 2\pi R^{2}\Sigma_{\rm disk}=
       4.4\times 10^{6}\,\alpha_{0.1}^{-4/5}M_{8}^{11/5}\mathdotM^{7/10}r_4^{5/4}\,\sunm$. 
For a conserved consideration 
of star formation efficiency of $\zeta=0.1$ in $10^{8}\,$yr \citep{Kennicutt1998}, $\zeta M_{\rm disk}$ 
will be converted into stars $M_{\star}$, and a fraction about $f_{*}=10^{2(1-n_{*})}\sim 0.1$ is
massive stars for star formation with a top-heavy IMF (the maximum and minimum masses are $(100,1)\sunm$, 
$n_{*}=0.5$ much more flatten than the Salpeter $n_{*}=2.35$), where initial mass function 
$dN_{*}\propto m_{*}^{-n_{*}}$. According to stellar evolution theory 
\citep[e.g., see Figure 16 in][]{Woosley2002}, about 10\% stellar mass is converted into BHs (even 
high fraction to BHs) for $\langle m_{*}\rangle\sim 100\sunm$ stars (evolved within $t_{\rm AGN}$). 
The black hole numbers are about 
$N_{\bullet}\approx 4.4\times 10^{2}\,\zeta_{0.1}f_{0.1}M_{\rm disk,6}/\langle m_{*}\rangle_{2}$, 
where $\langle m_{*}\rangle_{2}=\langle m_{*}\rangle/10^{2}\sunm$,
$\zeta_{0.1}=\zeta/0.1$ and $f_{0.1}=f_{*}/0.1$. Bondi explosion rates are about
\begin{equation}
\calR\approx\frac{N_{\bullet}}{t_{\rm rej}}
     \approx 1.6\,N_{\bullet,440}t_{\rm rej,270}^{-1}\,{\rm yr^{-1}},
\end{equation} 
where $N_{\bullet,440}=N_{\bullet}/440$ and $t_{\rm rej,270}=t_{\rm rej}/270\,$yr. The Bondi
explosion drives one transient appearance per year. Considering the uncertainties of $N_{\bullet}$, 
such an event rate is also comparable to AGN giant flare rates. It would be very interesting to 
observationally test the real AGN flares. Additionally, AMSs with neutron stars or $\sim 1\sunm$ 
black holes have fainter flares than that of $10\sunm$ black holes, however, their numbers could be 
more and thus explosion rates could be higher than a few per year generating relatively stationary 
emissions. To correctly answer this problem depends on more sophisticated studies of the IMF and 
stellar evolution in the SMBH-disks.

The Caltech-NRAO Stripe 82 Survey (CNSS), a dedicated radio transient survey \citep{Moorley2016} to
systematically explore the radio sky for slow transient phenomena on timescales between 
one day and several years, is very useful to detect the Bondi explosions suggested in this paper. 
The Very Large Array Sky Survey (VLASS) and Faint Images of the Radio Sky at Twenty (FIRST) cm 
survey, which have detected some variable objects \citep{Nyland2020}, are also very useful to test
the present scenario. Moreover, the {\it Fermi-}LAT monitoring campaign of non-blazar AGNs also 
provides unique opportunities for testing the Bondi explosion mechanism proposed here. Indeed, 
several radio-quiet AGNs clearly show variability at $>100\,$MeV \citep{Sahakyan2018}. Future 
joint radio and $\gamma$-ray observations of radio-quiet quasars have the potential to discover 
high-mass stellar black holes in SMBH-disks.

\subsection{Broad emission lines}
Bondi explosions couldn't affect the broad-line region, which is composed of discrete clouds keeping 
a pressure balance with its surrounding medium. The shocked clouds will reach temperatures up to 
$\sim 10^{8}\,$K, but its free-free cooling timescale is about
$t_{\rm ff}({\rm BLR})\approx 10^{5}\,T_{8}^{1/2}n_{10}^{-1}\,$s, where 
$n_{10}=n_{\rm cloud}/10^{10}\,{\rm cm^{-3}}$ is the cloud density in the BLR. 
Since $t_{\rm ff}({\rm BLR})$ is much shorter than the timescale of the shock crossing the BLR 
($t_{\rm BLR}=R_{\rm BLR}/V_{\rm exp}\sim 10^{8-9}\,$s), the shocked clouds will recover rapidly. 
Thus, the BLR itself is not strongly affected by Bondi explosions.

\section{discussion and conclusions}
Compact objects evolved from massive stars remain inside the accretion disks of SMBHs in AGNs 
and quasars. They inevitably form AMSs in the extremely dense environment of SMBH-disks. We 
suggest that the AMS black holes are episodically accreting with hyper-Eddington rates 
($\sim 10^{9}L_{\rm Edd}/c^{2}$) and duty cycles of $\sim 10^{-5}$. The active episodes are 
quenched by outflows from the hyper-Eddington accretion at a radius of $\sim10^{5}$ gravitational 
radius of the AMS black holes. The episodic hyper-Eddington
accretion allows stellar mass black holes to grow to exceed the well-known limit of the pair 
instability of massive stars.
A powerful fireball produced by the outflows will drive an intense and fast expansion, 
which we call a Bondi explosion. Non-thermal flares from the Bondi explosion decay with a 
timescale of $t^{-6/5}$ for a few months, resulting in transient emission in the radio, 
optical, UV, soft X-ray, and GeV bands. With an occurrence rate of a few per year, Bondi 
explosions may contribute to the variable light curves of quasars.

Observational searches for AMSs should focus on slowly varying transients in radio-quiet AGNs 
as manifested in the radio, optical, UV, and $\gamma$-ray bands. Joint radio (e.g., CNSS and 
VLASS) and $\gamma$-ray (e.g., {\it Fermi}-LAT) observations would be promising. This Letter 
only outlines the fate of AMSs in SMBH-disks, leaving much room for future investigations. 
Episodes of the BH accretion and the details of Bondi explosions should be more carefully
studied through numerical simulations. The cavity of in the SMBH-disk formed by the Bondi 
explosion determines the rejuvenation of the AMS and the Bondi explosion rates. 
The non-thermal emission from the explosion should also be calculated more self-consistently.

\acknowledgments
The authors are grateful to an anonymous referee for a helpful report clarifying 
several points of the manuscript.
JMW especially thanks Bin Luo for useful information on CNSS results. Helpful discussions are 
acknowledged with Y.-R. Li and Y.-Y. Songsheng from IHEP AGN Group. JMW thanks the support by 
National Key R\&D Program of China through grant -2016YFA0400701 and -2020YFC2201400 by 
NSFC through grants NSFC-11991050, -11991054, -11833008, -11690024, and by grant 
No. QYZDJ-SSW-SLH007 and No.XDB23010400. LCH is supported by the National Key R\&D Program 
of China through grant 2016YFA0400702 and the NSFC through grants 11721303 and 11991052.

\end{document}